\documentclass[12pt,english]{article}
\date{}
\usepackage{xcolor}
\usepackage[utf8x]{inputenc}
\usepackage{amsmath}
\usepackage{amsfonts}
\usepackage{amssymb}
\usepackage{graphicx,color}
\usepackage{babel}
\usepackage{fontenc}
\usepackage{hyperref}

\newcommand{\bea}{\begin{eqnarray}}
\newcommand{\eea}{\end{eqnarray}}
\newcommand{\be}{\begin{equation}}
\newcommand{\ee}{\end{equation}}

\renewcommand{\a}{\alpha}
\renewcommand{\b}{\beta}
\renewcommand{\c}{\gamma}
\renewcommand{\d}{\delta}
\newcommand{\e}{\epsilon}
\newcommand{\dsl}{\pa \kern-0.5em /}

\newcommand{\half}{\frac{1}{2}}
\newcommand{\pa}{\partial}

\newcommand{\nn}{\nonumber\\}

\def\be{\begin{equation}}
\def\ee{\end{equation}}

\begin{document}

\title{{\bf{Holographic complexity of `black' non-susy $D3$-brane and the high temperature limit }}}

\author{
{\bf {\normalsize Sourav Karar}$^{a,b}$\thanks{sourav.karar91@gmail.com}},
{\bf {\normalsize Sunandan Gangopadhyay}$^{b}$\thanks{sunandan.gangopadhyay@gmail.com, sunandan.gangopadhyay@bose.res.in }},
{\bf {\normalsize A. S. Majumdar}$^{b}$\thanks{archan@bose.res.in}}\\
$^{a}$ {\normalsize Department of Physics, Government General Degree College, Muragachha 741154, Nadia, India}\\
$^{b}${\normalsize S.N. Bose National Centre for Basic Sciences, Block-JD, Sector III, Kolkata 700106, India}\\
\\[0.3cm]}

\maketitle
\begin{abstract}
\noindent The holographic complexity of a `black' non-susy $D3$-brane is computed.
The difference in the holographic complexity between this geometry in the Fefferman-Graham coordinates and that of the $AdS_5$ geometry is obtained for a strip type subsystem. This is then related to the changes in the energy and the entanglement entropy of the system. We next take the high temperature limit of the change in complexity and observe that it scales with the temperature in the same way as the holographic entanglement entropy.
The crossover of the holographic complexity to its corresponding thermal counterpart is similar to the corresponding crossover of the holographic entanglement entropy in the high temperature limit.  
We further repeat the analysis for $\mathcal{N} =4$ super Yang-Mills theory and observe a similar behaviour.

\end{abstract}

\newpage

\section{Introduction}  
The computation of entanglement entropy (EE) in conformal field theories \cite{Bombelli:1986rw}-\cite{Eisert:2008ur} from a bulk gravity theory (which is asymptotically $AdS$) using the $AdS/CFT$ duality \cite{Maldacena:1997re, Aharony:1999ti} has been an area of active research in recent times. The computation of EE holographically is carried out using a formula 
given in \cite{Ryu:2006bv, Ryu:2006ef}. The formula enunciates that the holographic EE (HEE) of a subsystem $A$ in the gravity dual reads
\be
S_E = \frac{\mbox{Area} (\c_A^{min})}{4G_{(N)}}
\ee
where $\c_A^{min}$ is the $d$-dimensional minimal area surface in AdS$_{d+2}$ spacetime, the boundary of which matches with the
boundary of the subsystem $A$ and $G_{(N)}$ is the $(d+2)$-dimensional Newton's constant.

Interestingly it has been shown that the HEE for a very small subsystem satisfies a relation which looks similar to the first law of thermodynamics when
the system is excited \cite{Bhattacharya:2012mi, Allahbakhshi:2013rda}. A similar result have also been obtained in \cite{skarar} where a $(2+1)$-dimensional quantum many body system (exhibiting a Lifshitz symmetry \cite{ross} near their quantum critical point) is described by a $(3+1)$-dimensional dual gravity theory with a negative cosmological constant together with a massive vector field. It was shown that there exists an additional term in the relation analogous to the first law of thermodynamics which owed its origin to the presence of the massive gauge field.

Quantum complexity is yet another important quantity in the field of quantum information which has provided deep insights in the understanding of the properties of horizons of black holes. It has been realised that there may be a possible relation between complexity and fidelity susceptibility between two states appearing in the literature of quantum information \cite{Mohsen,Alishahiha:2017cuk}. Based on the prescription provided to compute the EE holographically, one may define holographic complexity (HC) following the proposal in \cite{sus1, sus2}
\be
C_V = \frac{V(\c)}{8\pi R G_{(N)}}
\label{c00}
\ee
where $R$ is the radius of curvature of the spacetime and $V(\c)$ is the volume of the part in the bulk geometry enclosed by the minimal hyper surface 
involved in the calculation of the HEE. The above proposal for evaluating the HC
by computing the volume enclosed by the minimal hypersurface gives the HC of a subregion and is known as holographic subregion complexity.
It should be noted however, that the change in HC
\be
\Delta C_V = C(\gamma_E)-C(\gamma_P)
\label{cha}
\ee
where $\gamma_P$ is the minimal hypersurface for the pure AdS$_5$ geometry and
$\gamma_E$ is the minimal hypersurface for the excited spacetime geometry, is physically important because it is a finite quantity although both 
$C(\gamma_E)$ and $C(\gamma_P)$ are divergent quantities.

We would like to point out that an alternative way of computing the HC of a system is from the bulk action evaluated
on the Wheeler-DeWitt patch \cite{sussk1}, \cite{sussk2}
\begin{eqnarray}
C_{W} =\frac{A(W)}{\pi \hbar}
\label{WDW}
\end{eqnarray}
where $A(W)$ is the action evaluated on the Wheeler-DeWitt
patch $W$ with a suitable boundary time. There are alternative recent proposals as well which relate bulk volumes to Fisher information \cite{fis} and fidelity susceptibility \cite{susc1}-\cite{susc3}. In a very recent paper \cite{erd}, a slightly different quantity has been proposed. The subregion complexity 
has been defined to be the integral of the scalar curvature over the region containing the regularized volume.

Computation of holographic subregion complexity have also been made recently in \cite{Karar:2017org} in the case of a $(3+1)$-dimensional
Lifshitz spacetime and the change in complexity has been related to the change in energy, change in EE and change in entanglement chemical potential
of the system.

In \cite{Bhattacharya:2017gzt}-\cite{Chakraborty:2017wdh}, a `black' non-susy $D3$-brane has been considered. This solution is known to have a decoupling limit \cite{Nayek:2016hsi} which in turn leads to a gravity dual picture.
The gravity dual description is exploited to calculate the EE of the quantum field theory associated with it. The asymptotically $AdS_5$ geometry of the `black' non-susy $D3$-brane in the decoupling limit allows one to write the HEE as a sum of two parts, namely, the HEE associated with the pure $AdS_5$ piece plus an additional piece which can be interpreted as the HEE of an excited state. The boundary stress tensor was next identified \cite{Balasubramanian:1999re, deHaro:2000vlm} using which it was shown that the HEE of the excited state 
satisfied the relation similar in form to the first law of thermodynamics. It was further demonstrated that at high temperature, the HEE makes a transition to the thermal entropy of the usual `black' $D3$-brane.

Recently, there has been a study investigating the relation between the HC, HEE and fidelity susceptibility for a spherical shell of $D3$-branes \cite{mom}. This provides a motivation to carry out the same investigation for the `black' non-susy $D3$-brane. In this paper, we obtain the change in subregion holographic complexity between the `black' non-susy $D3$-brane in the Fefferman-Graham coordinates and that of the $AdS_5$ geometry for a strip type subsystem. We then relate it to the change in HEE and the entanglement temperature. The high temperature limit of this is taken and is found to scale with the temperature in the same way as the HEE does thereby indicating a similar crossover of HC to its corresponding thermal counterpart. The analysis is repeated for $\mathcal{N} =4$ super Yang-Mills theory 
\cite{Witten:1998qj, Witten:1998zw} and a similar behaviour is obtained.

The paper is organised as follows. In section 2, we discuss  the holographic complexity for `black' non-susy $D3$-brane in the Fefferman-Graham coordinates. In section 3, we discuss
the high temperature limit of the complexity of the `black' non-susy $D3$-brane and $\mathcal{N} =4$ super Yang-Mills theory. We conclude in section 4.

\vspace{.5cm}


\section{Holographic complexity for `black' non-susy $D3$-brane} 

In this section, we shall first fix our notations by reviewing the computation of HC for a strip in AdS$_5$ space. The AdS$_5$ metric can be written
in Poincar\'e  coordinates as follows
\be\label{adsmetric}
ds^{2}= \frac{r_1^2}{z^2} \left( -dt^2+ \sum_{i=1}^{3}(dx^i)^2 +dz^2 \right)
\ee
where $r_1$ is the radius of curvature of the AdS space. We shall now compute the HC by calculating the volume
enclosed by the minimal Ryu-Takayanagi (RT) surface \cite{Ryu:2006bv, Ryu:2006ef}. We consider the entangling region in the boundary to be  a straight belt with 
width $\ell$ such that $-\frac{\ell}{2} \leq x^1 \leq \frac{\ell}{2}$ and $0 \leq x^2 ,x^3 \leq L$, where $L$ is 
the extent of the subsystem in the other spatial direction. We parametrize the extremal surface by $x^1 = x^1(z)$.
With this setup, the volume enclosed by the RT minimal surface reads \cite{sus1, sus2}
\be{\label{vol1}}
V^{(0)} = 2r_1^4 L^2 \int_{\e}^{z_t^{(0)}}  dz \; \frac{1}{z^4} \; x_1(z)
\ee
where $z_t^{(0)}$ is the turning point at which $dz/dx^1$ vanishes and $\e$ is the cut-off introduced to prevent
the integral from diverging at $z=0$. The profile of the minimal surface $x^1(z)$ by 
minimizing the RT area functional considering $z=z(x^1)$. The RT area functional for the metric (\ref{adsmetric}) is given by
\bea
A^{(0)} &=& \int_0^L dx_2 \; \int_0^L dx_3 \int_{-\frac{\ell}{2}}^{\frac{\ell}{2}}  dx_1 \frac{r_1^3}{z^3} \sqrt{1+z'(x_1)^2} \nn
 &=&  2 L^2 \int_{0}^{\frac{\ell}{2}}  dx_1 \frac{r_1^3}{z^3} \sqrt{1+z'(x^1)^2}
 \eea
 where $\prime$ denotes the derivative with respect to $x^1$. By minimizing this area functional we obtain
 $\frac{dz}{dx^1} =\sqrt{\left( \frac{z_t^{(0)}}{z}\right)^6 -1}$, which in turn gives the minimal surface profile and the width of the strip as follows
 \be\label{prfl1}
  x_1(z)=\int_z^{z_t^{(0)}} dz \frac{1}{\sqrt{\left( \frac{z_t^{(0)}}{z}\right)^6 -1}},\quad \quad 
 \frac{\ell}{2} = \frac{\sqrt{\pi}\Gamma\left(\frac{2}{3}\right)}{\Gamma\left(\frac{1}{6}\right)} z_{t}^{(0)} .
 \ee
 Using the above expression for $x_1(z)$ in eq.\eqref{vol1}, we obtain
 \bea{\label{unpertvol}}
 V^{(0)} &=& 2 L^2 r_1^4 \int_{\e}^{z_t^{(0)}} dz \; \frac{1}{z^4} \int_{z}^{z_t^{(0)}} dy \frac{1}{\sqrt{\left( \frac{z_t^{(0)}}{y}\right)^6 -1}} \nn
 &=& \frac{L^2 r_1^4 \ell}{3\e^3  } -\frac{4 \pi^{\frac{3}{2}}\Gamma\left(\frac{2}{3}\right)}{9\Gamma\left(\frac{1}{6}\right)} \frac{L^2 r_1^4}{\ell^2} ~.
 \eea
 Therefore, the HC for the $AdS_5$ geometry reads \cite{Ben-Ami:2016qex, Roy:2017kha}
 \bea\label{unpertcomp}
 \mathcal{C}_V^{(0)} &=& \frac{V^{(0)} }{8\pi r_1 G_{(5)}} \nn
 &=& \frac{L^2 r_1^3 \ell}{24 \pi G_{(5)} \e^3} - \frac{\sqrt{\pi} \Gamma\left(\frac{2}{3}\right) L^2 r_1^3}{18G_{(5)} \Gamma\left(\frac{1}{6}\right) \ell^2} ~.
 \eea 
With this result in hand, we now look at the `black' non-susy $D3$-brane solution of type IIB string theory in the Einstein frame. The metric of this 
 takes the form \cite{Bhattacharya:2017gzt}-\cite{Nayek:2016hsi}
 \bea\label{nonsusyd3}
& &  ds^{2}= F_1(r)^{-\frac{1}{2}}G(r)^{-\frac{\delta_{2}}{8}}\left[-G(r)^{\frac{\delta_{2}}{2}}dt^{2}+ \sum_{i=1}^{3}(dx^{i})^{2}\right]+F_1(r)^{\frac{1}{2}}
G(r)^{\frac{1}{4}}\left[\frac{dr^{2}}{G(r)}+r^{2}d\Omega_{5}^{2}\right]\nn
& & e^{2\phi} = G(r)^{-\frac{3\d_2}{2} + \frac{7\d_1}{4}}, \qquad
 \qquad F_{[5]} = \frac{1}{\sqrt{2}}(1+\ast) Q {\rm Vol}(\Omega_5)
\eea
where the functions $G(r)$ and $F(r)$ are defined as
\be\label{functions}
G(r)=1+\frac{r_{0}^{4}}{r^{4}},\qquad
F_1(r)=G(r)^{\frac{\alpha_1}{2}}\cosh^{2}\theta - G(r)^{-\frac{\beta_1}{2}}\sinh^{2}\theta ~.
\ee
Here $\d_1$, $\d_2$, $\a_1$, $\b_1$, $\theta$, $r_0$, $Q$ are the parameters characterizing the solution.
It is to be noted that the parameters are not all independent but they satisfy the following relations
\bea\label{relations}
& & \a_1 - \b_1=  0\nn
& & \a_1 + \b_1 =  \sqrt{10 - \frac{21}{2} \d_2^2 - \frac{49}{2} \d_1^2 + 21 \d_2\d_1}\nn
& & Q = (\a_1+\b_1)r_0^4 \sinh2\theta ~.
\eea
For $\d_2 =-2$ and $\d_1 =-\frac{12}{7}$, the solution \eqref{nonsusyd3} reduces to black $D3$-brane solution.
From now on we will put $\a_1+\b_1 = 2$ for simplicity. Therefore, from the first relation in eq.\eqref{relations}, we have $\a_1=1$ and $\b_1=1$.
In this case, the parameters $\d_1$ and $\d_2$ will be related (see the second equation in eq.\eqref{relations}) by
\be\label{deltareln}
42 \d_2^2 + 49 \d_1^2 - 84 \d_1\d_2 = 24~.
\ee
The function $F_1(r)$ given in eq.\eqref{functions} then reduces to 
\be\label{fonerho}
F_1(r) = G(r)^{-\frac{1}{2}} H(r)
\ee
where
\be\label{hr}
H(r) = 1 + \frac{r_0^4 \cosh^2\theta}{r^4} \equiv 1 + \frac{r_1^4}{r^4}~.
\ee
Using this form of $F_1(r)$ in eq.\eqref{nonsusyd3} leads to the metric in the Einstein frame to be
\be\label{metric1}
ds^{2}= H(r)^{-\frac{1}{2}}G(r)^{\frac{1}{4}-\frac{\delta_{2}}{8}}\left[-G(r)^{\frac{\delta_{2}}{2}}dt^{2}+ \sum_{i=1}^{3}(dx^{i})^{2}\right]+H(r)^{\frac{1}{2}}
\left[\frac{dr^{2}}{G(r)}+r^{2}d\Omega_{5}^{2}\right] 
\ee
where $H(r)$ is given in \eqref{hr}. It can be shown that the above reduces to \cite{Nayek:2016hsi}
\be\label{metric2}
ds^{2}= \frac{r^2}{r_1^2}G(r)^{\frac{1}{4}-\frac{\delta_{2}}{8}}\left[-G(r)^{\frac{\delta_{2}}{2}}dt^{2}+ \sum_{i=1}^{3}(dx^{i})^{2}\right]+
\frac{r_1^2}{r^2}\frac{dr^{2}}{G(r)}+r_1^{2}d\Omega_{5}^{2} 
\ee
where $r_1 = r_0 \cosh^{\half}\theta$ is the radius of the transverse 5-sphere which decouples from the five dimensional asymptotically $AdS_5$
geometry. This is possible if one looks into the region 
\be\label{zoom}
r \sim r_0 \ll r_0 \cosh^{\half}\theta
\ee
where $\theta \to \infty$ and the function $H(r) \approx r_1^4/r^4$.
In Fefferman-Graham coordinates \cite{fefferman}, the asymptotic limit of this metric takes the form \cite{Bhattacharya:2017gzt}
\be\label{metric3}
ds^{2}=\frac{r_1^{2}}{z^2}\left[-\left(1+\frac{3\delta_{2}}{8}\frac{z^4}{z_0^4}\right)dt^{2}+\left(1-\frac{\delta_{2}}{8}
\frac{z^4}{z_0^4}\right)\sum_{i=1}^{3}
(dx^{i})^{2} + dz^{2}\right]
\ee 
where $z_0^4 = r_1^8/r_0^4$. It is to be noted that the above geometry is valid near the boundary since weakly excited states have been considered in obtaining the above form of the metric. Furthermore, for $\delta_2=-2$, the above metric reduces to 
\be\label{metric3zz}
ds^{2}=\frac{r_1^{2}}{z^2}\left[-\left(1-\frac{3}{4}\frac{z^4}{z_0^4}\right)dt^{2}+\left(1+\frac{1}{4}
\frac{z^4}{z_0^4}\right)\sum_{i=1}^{3}
(dx^{i})^{2} + dz^{2}\right].
\ee 
It is evident that the above metric does not match with the $AdS_5$ black hole metric \cite{Mohsen}
\be\label{metric3zzz}
ds^{2}=\frac{r_1^{2}}{z^2}\left[-h(z)dt^{2}+\frac{1}{h(z)}dz^2 +\sum_{i=1}^{3}
(dx^{i})^{2} \right]~;~h(z)=1-mz^4~,~m=\frac{1}{z_{0}^4}
\ee 
with $m$ being a constant. Near the boundary this takes the form
\be\label{metric3Zqq}
ds^{2}=\frac{r_1^{2}}{z^2}\left[-(1-mz^4)dt^{2}+(1+mz^4)dz^2 +\sum_{i=1}^{3}
(dx^{i})^{2} \right].
\ee 
It should be noted that there are differences between the forms of the metrics (\ref{metric3zz}) and (\ref{metric3Zqq}) in terms of the numerical factors of the metric coefficients in eq.(s)(\ref{metric3zz}) and (\ref{metric3Zqq}). Indeed, the factor in eq.(\ref{metric3zz}) is in front of
$(dx^{i})^{2}$ whereas in eq.(\ref{metric3Zqq}) is in front of $dz^2$.
We shall refer to these results in the subsequent discussion.

\noindent With this set up in place, we are now ready to calculate the HC for `black' non-susy $D3$-brane. It should be remembered that
we are interested in the change in complexity due to small perturbation in the background metric.
To see the change in complexity due to change in metric perturbation, we keep the length $\ell$ fixed. Under this condition
the expression for volume enclosed by the RT minimal surface for the above metric \eqref{metric2} for the same entangling region reads 
\bea{\label{pertvol}}
V &=& 2 L^2 r_1^4 \int_{\e}^{z_t^{(0)}} dz \; \frac{1}{z^4}\left( 1-\frac{\delta_2}{8}\frac{z^4}{z_0^4}\right)^{3/2} \int_{z}^{z_t^{(0)}} dy \frac{1}{\sqrt{\left( \frac{z_t^{(0)}}{y}\right)^6 -1}} ~.
\eea
Now as we are interested in the region near the boundary, we keep only the first order terms in the perturbation to write eq.\eqref{pertvol}
as
\bea
V &=& 2 L^2 r_1^4 \int_{\e}^{z_t^{(0)}} dz \; \frac{1}{z^4}\left( 1-\frac{3\delta_2}{16}\frac{z^4}{z_0^4}\right) \int_{z}^{z_t^{(0)}} dy \frac{1}{\sqrt{\left( \frac{z_t^{(0)}}{y}\right)^6 -1}} \nn
&=& V^{(0)}  - \frac{3\delta_2}{8} L^2 r_1^4  \int_{\e}^{z_t^{(0)}} dz \; \frac{1}{z_0^4} \; \int_{z}^{z_t^{(0)}} dy \frac{1}{\sqrt{\left( \frac{z_t^{(0)}}{y}\right)^6 -1}}
\eea
where V$^{(0)}$ is the volume of the $AdS_5$ background. Hence the change in the HC due to a small perturbation in the background metric \eqref{metric2} in terms of change in volume $\Delta V$ is given by \cite{sus1, sus2}
\bea\label{comp}
\Delta\mathcal{C} &=& \frac{\Delta V}{8\pi r_1 G_{(5)}} \nn
&=& -\frac{3\delta_2}{512 \pi^{\frac{3}{2}} G_{(5)}} \frac{\Gamma\left(\frac{1}{6}\right)^2 \Gamma \left(\frac{5}{6}\right) L^2 r_1^3 \ell^2}
		{\Gamma\left(\frac{2}{3}\right)^2 \Gamma\left(\frac{1}{3}\right)z_0^4}~.
\eea
For $\delta_2=-2$, the above result simplifies to
\bea\label{simpli}
\Delta\mathcal{C} = \frac{3}{256 \pi^{\frac{3}{2}} G_{(5)}} \frac{\Gamma\left(\frac{1}{6}\right)^2 \Gamma \left(\frac{5}{6}\right) L^2 r_1^3 \ell^2}
		{\Gamma\left(\frac{2}{3}\right)^2 \Gamma\left(\frac{1}{3}\right)z_0^4}~.
\eea
This agrees in form with the change in complexity for the $AdS_5$ black hole metric near the boundary (\ref{metric3Zqq}). The change in complexity in this case reads
\bea\label{simpli100}
\Delta\mathcal{C} = \frac{4m}{256 \pi^{\frac{3}{2}} G_{(5)}} \frac{\Gamma\left(\frac{1}{6}\right)^2 \Gamma \left(\frac{5}{6}\right) L^2 r_1^3 \ell^2}
		{\Gamma\left(\frac{2}{3}\right)^2 \Gamma\left(\frac{1}{3}\right)}~,~m=\frac{1}{z_{0}^4}.
\eea
We make an important observation now. We find that the change in subregion holographic complexity (\ref{comp}) for the `black' non-susy $D3$-brane in the Fefferman-Graham coordinates
(\ref{metric3}) does not reduce to the change in subregion holographic complexity for the $AdS_{5}$ black hole (\ref{simpli100}) in the $\delta_{2}=-2$ limit as can be easily seen from eq.(\ref{simpli}). It is reassuring to note that this observation has  been confirmed in a recent paper for both strip and spherical subsystems \cite{shibu}\footnote{This paper appeared in the arXiv 3 months after our paper.}. It is easy to see that there is an error of 25$\%$ in the result (\ref{simpli}) from the actual result (\ref{simpli100}) for the $AdS_5$ black hole. The reason for this slight mismatch is evident. The asymptotic form of the `black' non-susy $D3$-brane metric in the Fefferman-Graham coordinates (\ref{metric3}) given in \cite{Bhattacharya:2017gzt} does not reduce to the $AdS_{5}$ black hole metric in the $\delta_{2}=-2$ limit. In this context we would like to mention that in \cite{shibu}, it has been stated that although the change in holographic complexity for the `black' non-susy $D3$-brane computed in the Fefferman-Graham coordinates does not match with the change in holographic complexity of the $AdS_{5}$ black hole metric in the $\delta_{2}=-2$ limit, the change in holographic entanglement entropy matches. We would like to point out that this agreement should not be taken seriously as the asymptotic form of the `black' non-susy $D3$-brane metric in the Fefferman-Graham coordinates (\ref{metric3}) does not reduce to the $AdS_{5}$ black hole metric (\ref{metric3zzz}) in the $\delta_{2}=-2$ limit and an agreement of the holographic entanglement entropy is merely a coincidence. The reason for this is clear as the recasting of the `black' non-susy $D3$-brane metric in the Fefferman-Graham coordinates has been carried out after making certain approximations and then taking the asymptotic limit of the metric (\ref{metric2}).
 
\noindent We now proceed to study the relation between the change in complexity with the stress energy tensor.
The stress energy tensor for an asymptotically local $AdS$ metric \cite{fefferman}
 \be\label{metric}
ds_{d+1}^2=\frac{r_1^2}{z^2}\bigg(dz^2+g_{\mu\nu} dx^\mu dx^\nu\bigg),
\ee where $g_{\mu\nu}=\eta_{\mu\nu}+h_{\mu\nu}(x,z)$ with \be
h_{\mu\nu}(x,z)=h^{(0)}_{\mu\nu}(x)+h^{(2)}_{\mu\nu}(x)z^2+\cdots+z^d\left(h^{(d)}_{\mu\nu}(x)
+{\hat h}^{(d)}_{\mu\nu}(x) \log z\right)+\cdots \ee 
reads
\be
 \langle T_{\mu\nu}\rangle
=\frac{d \;r_1^{d-1}}{16\pi G_{(N)}}\;h_{\mu\nu}^{(d)}~.
\ee
One can easily see that for the metric \eqref{metric3} the expectation values of the stress energy tensor reads \cite{Balasubramanian:1999re}
\bea\label{ourstensor}
\langle T_{tt}\rangle &=& \frac{-3 r_1^3 \d_2}{32\pi G_{(5)} z_0^4} \\
\langle T_{x_i x_j}\rangle &=& \frac{- \rho_1^3 \d_2}{32\pi G_{(5)}z_0^4} \delta_{ij}
\eea
where $i,\,j=1,\,2,\,3$.
Now using the fact that the total energy can be recast as 
\be
\Delta E =\int_0^L dx_2 \int_0^L dx_3 \int_{-\frac{\ell}{2}}^{\frac{\ell}{2} } \langle T_{tt}\rangle = L^2 \ell \langle T_{tt}\rangle
\ee
the expression for change in complexity (\ref{comp}) can be restated as follows
\be\label{compenergy}
\Delta\mathcal{C} = \frac{3 \Gamma \left( \frac{5}{6}\right)^2}{2\pi \Gamma \left(\frac{1}{3}\right)^2} \frac{\Delta E}{T_{ent}}
\ee
where $T_{ent}$ is the entanglement temperature \cite{Allahbakhshi:2013rda, Bhattacharya:2017gzt}
\be
T_{E}=\frac{24 \Gamma\left(\frac{5}{6}\right)\Gamma^{2}\left(\frac{2}{3}\right)}{\sqrt{\pi}\Gamma\left(\frac{1}{3}\right)
\Gamma^{2}\left(\frac{1}{6}\right)}\frac{1}{\ell}~.
\ee
Again we know that the decoupled theory of `black' non-susy $D3$-brane satisfies the first law of entanglement thermodynamics \cite{Bhattacharya:2017gzt}
\be\label{law}
\Delta E = T_{E}\Delta S_{E} + \frac{3}{5} \Delta P_{x_{1}x_{1}} V_{3}
\ee
where $V_{3}= L^2 \ell$ is the volume of the subspace and $\Delta P_{x_1 x_1} = \langle T_{x_1 x_1} \rangle$ is the
entanglement pressure. We can use this first law to recast eq.(\ref{compenergy}) as
\be\label{compentropy}
\Delta\mathcal{C} = k_1 \Delta S_{E} + k_2 \frac{\Delta P_{x_1 x_1} V_3}{T_{ent}}
\ee
where the constants $k_1$ and $k_2$ are given by
\begin{eqnarray}\label{const1}
k_1 &=& \frac{3}{2\pi}\frac{\Gamma^2 (5/6)}{\Gamma^2 (1/3)}\\
\label{const2}
k_2 &=& \frac{3}{5}k_1~.
\end{eqnarray}
Eq.(\ref{compentropy}) relates the change in HC with the change in HEE and the entanglement pressure. The relation can be considered to be an analogous expression corresponding to the relation for the change in HEE regarded as the first law of entanglement thermodynamics. Furthermore, the relation also leads to upper or lower bounds for the HC depending on the sign of the entanglement pressure.

\section{High temperature limit}
In this section, we look at the high temperature limit of the finite part of complexity for `black' non-susy $D3$-brane. 
Setting $z/z_t =y$, the entangling region $\ell$ can be written as
\bea\label{l}
\frac{\ell}{2} &=& z_t \int_0^1 dy \frac{y^3\left(1-\frac{3\d_2}{16}\frac{z_t^4}{z_0^4}\right)}{\sqrt{\left(1-\frac{\d_2}{8}\frac{z_t^4}{z_0^4} y^4\right)
\left[1 - y^6 -\frac{3\d_2}{8}\frac{z_t^4}{z_0^4} y^4\left(1-y^2\right)\right]}}\nn
& \equiv & z_t {\cal I}\left(\frac{z_t}{z_0}\right) ~. 
\eea
The total volume is given by
\be
V= 2L^2 \int_{\e}^{z_t} dz\; \frac{r_1^4}{z^4} \left(1- \frac{\d_2 z^4}{8z_0^4}\right)^{3/2} \;
	\int_{z}^{z_t} dz_1 \; \frac{\sqrt{1-\frac{3\d_2z_t^4}{8z_0^4}}}
	{\sqrt{\left(1- \frac{\d_2 z^4}{8z_0^4}\right)\; \left[ \frac{z_t^6}{z_1^6}-1 -\frac{3\d_2 z_t^4}{8z_0^4}\left(\frac{z_t^2}{z_1^2}-1\right)\right]}}~.
\ee
This expression for volume is divergent and we are interested only in the behaviour of the finite part of the volume. The finite
part of the volume in this case is given by
\be
V_{finite} = \frac{2L^2 r_1^4}{3z_t^2}
				\int_0^1 dy\; \frac{{}_2F_1(-0.75,-1.5,0.25,-\frac{\d_2 z_t^4 y^4}{8z_0^4})\sqrt{1-\frac{3\d_2z_t^4}{8z_0^4}}}
				{y^3 \sqrt{\left(1-\frac{3\d_2z_t^4 y^4}{8z_0^4} \right) \left[1-y^6 -\frac{3\d_2 z_t^4}{8z_0^4} x^4(1-x^2)\right]}}
\ee
where ${}_2F_1$ is the hypergeometric function.
Now in the high temperature limit $z_t \rightarrow z_0$, using eq.\eqref{l} one can recast $\mathcal{C}_{finite}$ as
\be\label{compt}
\mathcal{C}_{finite} = \frac{V_{finite}}{8\pi r_1 G_{(5)}} \sim \frac{{}_2F_1(-0.75,-1.5,0.25,-\frac{\d_2 }{8})L^2 l r_1^3 }{G_{(5)} z_t^3 }~.
\ee
Now using the five dimensional Newton's constant $G_{(5)} = (\pi r_1^3)/(2 N^2)$, where $N$ is the number of branes and $1/(\pi z_0) = T$,
where $T$ is the temperature of the standard black $D3$-brane, we get from \eqref{compt} 
\be
\frac{\mathcal{C}_{finite}}{V_3} = {}_2F_1(-0.75,-1.5,0.25,-\frac{\d_2 }{8})\frac{\pi}{12}  N^2 T^3~.
\ee
To conclude this section, we look at the high temperature behaviour of the $AdS$ black hole geometry in this case dual to the familiar $\mathcal{N} =4$ super
Yang-Mills theory. The $AdS$ black hole geometry is given by \cite{Witten:1998qj, Witten:1998zw}
\be\label{metricyang}
ds^2 = r_1^2 \left[ \frac{du^2}{hu^2} +u^2 (-h dt^2 +dx_1^2 +dx_2^2 +dx_3^2 +d\Omega_5^2)\right]
\ee
with $h=1-\frac{u_0^4}{u^4}$ and $u_0 = \pi T$, where $T$ is the temperature of the black hole.
We now compute the complexity for a straight belt with 
width $\ell$( $-\frac{\ell}{2} \leq x_1 \leq \frac{\ell}{2}$ ; $0 \leq x_2 ,x_3 \leq L$). The length of the strip in this case is
\be\label{l1}
\frac{\ell}{2} = \frac{1}{u_t} \int_1^{\infty} dy \; \frac{1}{\sqrt{\left(y^4 - \frac{u_0^4}{u_t^4} \right)\left(y^6 - 1\right)}}~.
\ee
The total volume enclosed by the Ryu-Takayanagi surface is
\be
V= 2L^2 r_1^4 \int_{u_t}^{\infty} du\;\frac{u^4}{\sqrt{u^4 -u_0^4}} \; \int_{u_t}^{u} du_1 \frac{1}{\sqrt{\left(u_1^4 -u_0^4\right)\left(\frac{u_1^6}{u_t^6}-1\right)}}~.
\ee 
The finite part of this volume is given by 
\be
V_{finite} = \frac{2}{3}L^2 r_1^4 u_0^2 \int_1^{\infty} dy \frac{{}_2F_1(-0.75,0.5,0.25,\frac{u_0^4}{y^4 u_t^4})y^3 }
		{\sqrt{\left(y^4 -\frac{u_0^4}{u_t^4}\right)\left(y^6 - 1\right)}}~.
\ee
Now in the high temperature limit $u_t \rightarrow u_0$, using eq.\eqref{l1} we can compute the volume to be
\be
V_{finite}= \frac{1}{3} {}_2F_1(-0.75,0.5,0.25,1) L^2 \ell r_1^4 u_0^3~.
\ee
Therefore, the finite part of the complexity for this case is given by
\be
 \frac{\mathcal{C}_{finite}}{V_3}= {}_2F_1(-0.75,0.5,0.25,1) \frac{\pi}{12}N^2 T^3~.
\ee
It is interesting to note that the high temperature limit of the complexity in both cases scales as $T^3$ which is the same scaling behaviour
which the HEE exhibits in 5 -dimensional bulk theory \cite{dong}.
\section{Conclusions} 
In this paper, we compute the holographic complexity of a `black' non-susy $D3$-brane. From this we obtain the difference in holographic complexity between this geometry in the Fefferman-Graham coordinates and the $AdS_5$ spacetime for a strip type subsystem. The difference is found to be a finite quantity. The finiteness of this expression is important in order to propose the holographic dual to Fisher information. Interestingly, we observe that this change in subregion holographic complexity does not reduce to the change in subregion holographic complexity for the $AdS_{5}$ black hole in the $\delta_{2}=-2$ limit and that there is an error of 25$\%$ from the actual result. This observation has  been confirmed in a recent paper \cite{shibu} (for both strip and spherical subsystems) which appeared in the arXiv 3 months after our paper. The reason for this disagreement owes its origin to the fact that the asymptotic form of the `black' non-susy $D3$-brane metric in the Fefferman-Graham coordinates does not reduce to the $AdS_{5}$ black hole metric in the $\delta_{2}=-2$ limit. 
We then expressed the change in the holographic complexity in terms of the changes in the energy and the entanglement entropy of the system. 
The motivation for carrying this out is to obtain an analogous expression corresponding to the first law of entanglement thermodynamics.
The high temperature limit of this is computed next. 
It is observed that it scales with the temperature in the same way as the holographic entanglement entropy does \cite{Bhattacharya:2017gzt} which in turn displays a similar crossover of holographic complexity to its corresponding thermal counterpart. The analysis is then carried out for $\mathcal{N} =4$ super Yang-Mills theory and a similar behaviour is obtained once again.
Further in depth analysis may be required to ascertain if this behaviour of similar crossover for holographic complexity and holographic entanglement entropy is a universal feature in $5$-dimensional spacetime geometries.
As a final remark, we would like to point out that the calculations carried out in this paper is for strip like subregions. Extending this work for spherical disk like subregion would be interesting and can be taken up as a future work.


\section*{Acknowledgements} S.G. acknowledges the support by DST SERB under Start Up Research Grant (Young Scientist), File No.YSS/2014/000180. S.G. also acknowledges the support of the Visiting Associateship programme of IUCAA, Pune.
The authors thank the referee for useful comments which has considerably improved the discussion in this paper.


\end{document}